\documentclass[aps,12pt,nofootinbib,showkeys]{revtex4}%
\usepackage{amsfonts}
\usepackage{amsmath}
\usepackage{amssymb}
\usepackage{graphicx}%
\providecommand{\U}[1]{\protect\rule{.1in}{.1in}}

\begin{document}
\preprint{ }
\title[ ]{The Schr\"{o}dinger Cat: Physics, Myth, and Philosophy}
\author{C. S. Unnikrishnan}
\affiliation{Tata Institute of Fundamental Research, Homi Bhabha Road, Mumbai 400005, India}

\begin{abstract}
The most discussed of ``live'' metaphors in physical sciences is that of the Schr\"odinger cat. Introduced in the early, but mature days of the new quantum theory, in 1935, the parable of the cat has provoked and enlivened debates on the meaning of the quantum theory, on the ontology of the quantum wavefunction, on the puzzle of the collapse of the wavefunction, on the meaning of quantum measurement, and on the boundary between the quantum world and the classical world. In this article, I will discuss these issues of the quantum theory focussing on various aspects of the metaphor of the Schr\"odinger cat.

\end{abstract}

\startpage{1}
\endpage{102}
\maketitle
\begin{quote}
	This is a slightly revised version of my article written in 2005, and published in the compiled volume, `History of Science and Philosophy of Science', p.~313-338 (Ed. P. K. Sengupta, Centre for Studies in Civilization, PHISPC, Pearson-Longman, Delhi, 2010). It was also published in \textit{Sandhaan}, Vol. V, No.2, 1-52 (2005), the journal of the Centre for the Studies in Civilizations, New Delhi). 
\end{quote}
\tableofcontents

\section*{Preface}
In 1935, Erwin Schr\"odinger was inspired to think deeply about the quantum mechanical description of two systems that interacted and then separated, by the discussion in the Einstein-Podolsky-Rosen (EPR) paper that questioned the completeness of standard quantum mechanics \citep{EPR1935}. In several papers he discussed the central and universal trait of `entanglement' in quantum mechanics \citep{Schrodinger1935,Schrodinger-Cat,SCat-Eng}. He realized that interactions that form correlations `entangle' the individual descriptions, resulting in only a joint and less detailed description of the physical state, and that the individual wavefunctions ceased to exist. He wrote,
\begin{quote}
	When two systems, of which we know the states by their respective representatives, enter into temporary physical interaction due to known forces between them, and when after a time of mutual influence the systems separate again, then they can no longer be described in the same way as before, viz. by endowing each of them with a representative of its own. I would not call that one but rather the characteristic trait of quantum mechanics, the one that enforces its entire departure from classical lines of thought. By the interaction the two representatives (or $\psi$-functions) have become entangled.
\end{quote}
An observation on one system then disentangles the description ($\psi$-function) and restores individual realities, in the quantum mechanical sense, in terms of individual wavefunctions. \emph{One of these systems could be macroscopic}, in terms of size, mass etc., and then new conceptual issues arise.  He discussed a theory of quantum measurement based on these insights. He demonstrated that the quantum uncertainty was not a blurring of physical variables, confined to the microscopic domain of atomic systems; the possibility of an interaction expanded the domain of uncertainty and fuzziness to the macro-world. In these studies arose the ever-living fable of the life of a cat that got entangled with the randomly timed radioactive decay of an atom. The graphic scenario exemplified since then the central tenets of the problem of `quantum measurement', as well as the troubling notion of a quantum-classical divide. The `Schr\"odinger cat' (also referred to as ``Schr\"odinger's cat'') challenged the universal applicability of quantum mechanics and its ability to represent physical states faithfully, further supporting Einstein's views that were expressed in the EPR paper. Indirectly, it raised new questions about the role of the conscious observer in quantum mechanics. With life mixed with physics and quantum mysteries, it generated large popular interest. Philosophers discussed the nature and meaning of physical reality in quantum mechanics and its relation to the observer, in the domain well beyond the microscopic world of quantum mechanics. 

There were also intellectual compromises: the superposition and interference of the quantum states of a single `large enough' system being confused as the realization of a quantum state of  the Schr\"odinger cat  is common. The perennial confusion between `a cat in a superposition of its non-commensurate physical states', and the very different idea of `an entangled state of a Schr\"odinger cat'  is serious and widespread. Some of these myths are here to stay, being ingrained in the collective psyche by repeated statements.

The foundational problems raised in the early years of quantum mechanics, especially the problem of quantum measurement, remain open. The aim of this article is to explore the genuine issues in physics and philosophy that are prompted by Schr\"odinger's entangled cat. 

\section{The Quantum Theory}

The years between 1924 and 1927 saw accelerated revolutionary changes in the understanding of the microscopic atomic physics. The old quantum theory that slowly developed in the work of Planck, Einstein, Bohr and others saw a sudden change in its physical and mathematical understanding and formalism. One line of thought starting with Louis de Broglie’s proposal that particles have wave nature matured in Erwin Schr\"odinger succeeding in writing a mathematically and physically consistent ``wave equation'' for the entity possessing the wave-particle duality \citep{Schrodinger1926}. With this equation, soon named the Schr\"odinger equation, accurate calculations of quantized energies and spectral lines in atoms could be made. Also, the equation enabled calculations of many other phenomena pertaining to microscopic physics. In a parallel development, Werner Heisenberg wrote down quantum rules for calculating these quantities based on a mathematical formalism in which physical quantities were represented by matrices – arrays of complex numbers. An important mathematical property of these matrices were that they were non-commuting; the order of their multiplication was important and reversing the order gave a different value for their product in general. This property accounted for the wave-particle duality in the Schr\"odinger formalism. Soon, Schr\"odinger showed that the two formalisms were equivalent, and the mathematical foundations of the new quantum theory were established. In 1927, the important work by Paul Dirac extended the quantum formalism~\citep{Dirac1927a} and linked it with the mathematics of classical mechanics, and he also wrote down the quantum wave equation for the electron, consistent with the theory of relativity.  Of crucial importance were the contributions by Max Born and John von Neumann, for the physical interpretation of the mathematical elements of the theory. 

One important aspect of the Heisenberg rules is that some combinations of physical quantities (`observables') cannot be determined accurately simultaneously. Multiple observations yield values with a statistical dispersion, and the product of the dispersion in the values of repeated measurements of two such observables on similar systems have a minimum universal value, given by the Planck's constant $\hbar$; this is called the uncertainty relation. For example, the position and the momentum of a particle are such a pair. It is important to note that the uncertainty relation in quantum mechanics is defined only for repeated measurements in a statistical ensemble, and never for a single particle or a physical system. On the other hand, the uncertainty relation can be applied to a single system for \emph{estimating the expected uncertainty in the physical quantities}, when the physical state is known.

In the present usage both the Schr\"odinger and Heisenberg approaches are practised, though the fundamental equation of quantum motion is usually written down in the Schr\"odinger form, as a differential equation. The Schr\"odinger equation describes how the wavefunction (or the $\psi$-function) evolves in time under the action of external interactions. The wavefunction itself is an unobservable, but all observables of the physical system are related to it by certain statistical relations. Usually, solving the equation for the specific problems gives rise to quantized discrete energy states, as well as some un-quantized states. The uncertainty relation follows in this approach from the properties of the wavefunction, and its relation to the various observables.

A crucial piece of interpretational discovery was made by M. Born regarding the Schr\"odinger wavefunction, and this discovery provided the most essential and at the same time most problematic aspect of the quantum theory; Born realized that the connection between the wavefunction and results of a measurement or observation was essentially statistical. Thus, the square of the wavefunction provides a probability density for various observables associated with the system. It became clear after Born’s work that quantum mechanics had no predictive power for single events. It is a theory of statistical predictions of the properties of a microscopic system, like an atom. This turn in the meaning of the theory, which had a deterministic differential equation as its basic equation, came as an unpleasant shock to the pioneers like Einstein and Schr\"odinger.

While it is possible to think of the wavefunction for a single particle as somehow a wave associated with the particle, as envisaged by de Broglie originally, this picture breaks down when one considers several particles. The wave that one is referring to is not a real wave in space, propagating as time progresses. It is a wave in the sense of an abstract mathematical function that admitted the properties of `superposition' and `interference', but one that merely represents the potentialities of measurement results that can be obtained from the physical system. In other words, the wavefunction itself is not an observable. Then `Born's rule' associates the square of the wavefunction to the probability of the realization of a particular result being obtained. The quantum theory has only statistical predictions and the wavefunction provides the mathematical link between the physical system and these probabilities. Right at this point it should have been clear that one should not physically associate the wavefunction and the quantum system, and \emph{one should not see the physical system or the particle as an entity possessing the dual property of a wave and the particle}.

\section{The Uncomfortable Quantum Rules}
The basic idea that is operative in the new mechanics is that the representations of two distinct physical states of a system can be added linearly in various ways to get a new valid physical state of the system. This never happens in the familiar world of particles around us. For example, a familiar system that exhibits just two discrete possible physical states is a coin that is tossed. It can either turn a `Head’ or a `Tail’. This happens randomly. But there is no concept of a physical state that is a sum or difference of these two states in a mathematically literal sense. However, when one considers a physical wave (on water for example) it makes perfect sense to think of the state of a new wave that is a sum or difference of two other waves – wherever the crests add together, the new wave just gets amplified and when a crest and trough coincide the result is the absence of a wave. So waves can cancel or augment each other, and their sum or difference is also a physical wave. But if a coin, the states of which are not waves, were to be described quantum mechanically, then a state that is a sum or difference of `Heads’ and `Tails’ is also valid and one needs to imagine what that might mean. At the same time, when an observation is made to determine the state of the coin, one and only one of these two possibilities turn up, and that too randomly. \emph{It is this conceptual wave-particle duality that quantum mechanics embodies, and this duality is unfamiliar to our experience, even though each aspect separately is very familiar}. It is this uncomfortable consequence of the supposed applicability of quantum mechanics to the everyday physical world that Schr\"odinger tried to express dramatically in his discussion of the quantum states of a cat. His analysis revealed an even more puzzling aspect of the superposition of the wavefunctions, which he called the `entanglement'.

Physicists who took the wave-particle duality of matter seriously in the sense of associating both a particle nature and also a wave nature to a single entity, thought of microscopic particles as some fuzzy (blurred, or distributed) entity. Schr\"odinger had spent a great deal of effort in trying to understand the quantum fuzziness and its implication to the larger world. Does the uncertain fuzziness suggested by the uncertainty principle and the mathematical shape of the wavefunction of an electron or an atom become sharply defined matter, without any fuzziness, in the macroscopic world? He discussed the following ``burlesque'' (translated as `ridiculous') example to bring out the strange implications associated with the mathematical machinery of the new quantum theory \citep{SCat-Eng}
\begin{quote}
	One can even set up quite ridiculous cases. A cat is penned up in a steel chamber, along with the following diabolical devise (which must be secured against direct interference by the cat): in a Geiger counter there is a tiny bit of radioactive substance, so small that \textit{perhaps} in the course of 1 hour one of the atoms decays, but also, with equal probability, perhaps none; if it happens, the counter tube discharges and through a relay releases a hammer which shatters a small flask of hydrocyanic acid. If one has left this entire system to itself for an hour, one would say that the cat is still alive if meanwhile no atom has decayed. The first atomic decay would have poisoned it. The wavefunction of the entire system would express this by having in it the living and the dead cat (pardon my expression) mixed or smeared out in equal parts. 
	
	It is typical of these cases that an indeterminacy originally restricted to the atomic domain becomes transformed into macroscopic indeterminacy, which can then be \textit{resolved} by direct observation. That prevents us from so naively accepting as valid a ``blurred model'' for representing reality.
\end{quote}
This example, now called `the Schr\"odinger cat', is the central theme of our discussion. Before we discuss the various aspects of the Schr\"odinger cat, it might be useful to have a brief look at the mathematical machinery that forms the basis of all the discussions related to this example. 

\section{The Mathematics and Physics of Linear Superposition}
The basic mathematical machinery of quantum mechanics is simpler than that of the older classical physics. The feature that is most important is that of the linear superposition, familiar from the mathematics of vectors and physics of waves. In quantum mechanics, the physical state of a particle has a meaning only after a measurement. But its mathematical state can be represented by a wavefunction, or more generally as a vector in an abstract space. By `vector' we mean an abstract entity with various components along different directions, representing physical quantities like velocity, force etc. The difference in quantum mechanics is that these components are complex numbers in general and the components do not refer to spatial directions; the `space' in which the vectors represent the physical system is an abstract mathematical space, which could even have  infinite dimensions. The fundamental premise of the quantum theory is that if A and B are two legitimate vectorial representations of possible physical states of a particle, then their linear addition, $aA+bB$,  is also a legitimate state of the particle. The vectors are added after arbitrarily stretching or contracting them by complex number scaling factors `a' and `b', with the restriction $|a|^2+|b|^2=1$. Then the Born rule says that a measurement will reveal the property A with the probability $|a|^2=aa^*$ and B with the probability $|b|^2$. 

\begin{figure}
	\centering
	\includegraphics[width=0.5\linewidth]{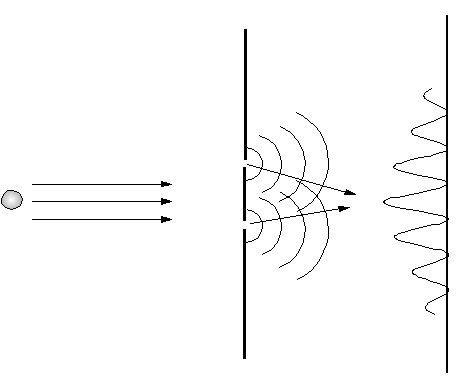}
	\caption{The double-slit experiment. An interference pattern (regions where there are particles separated by regions where there are no particles detected) is observed with light, atoms and electrons.  The natural explanation is in terms of waves that split and superpose. But the same physical behaviour is observed even when the particles are sent `one by one' forcing us to attribute a `wave' with even single particles.}
	\label{fig:fig1}
\end{figure}
First we mention a physical example that illustrates how superpositions become necessary. In the familiar example of a double-slit interference experiment, light falling on a screen with two slits forms an interference pattern with bright and dark fringes on a detector screen far away (Figure 1). If light is treated as waves, there is an immediate explanation in terms of the familiar interference patterns involving waves, like waves on water. It is possible to do the experiment such that the source of light is a single atom or a small number of atoms, and then light is received at the screens as discrete photons, much like particles arriving, well separated in time (as `seen' by an electronic  detector), and not as a continuous wave. Yet, the final pattern of interference obtainable on a long exposure camera is identical to the one obtained by the instantaneous wave description. \emph{The only logical possibility is to think of each photon being associated with a wave that interferes independent of when and where the next photon arrives}. In such a picture, the state of the photon after the slits has to be very different from our usual description of the state of a particle with a single unique trajectory. Somehow one has to account for the waves that will split and propagate through both the slits and then interfere even though we have only one indivisible particle to deal with (figure \ref{fig:fig1}). That is where the concept of superposition becomes necessary. The state of the photon is described as containing two parts, one corresponding to the propagation through the upper slit (U), and another through the lower slit (D), in equal proportions. These two parts can then interfere and represent the observed pattern. This does not imply that the particle itself splits at the slits and propagates in the two paths simultaneously – the associated waves do that job, if one needs to form a graphical representation. In fact, the particle cannot be in both slits in any manner without grossly violating all conservation principles in physics. So, the state of the particle after the slits is \emph{represented} as a wavefunction, 
\begin{equation}\label{key}
	|State\rangle\sim |Wave\mbox{-}U\rangle+|Wave\mbox{-}D\rangle
\end{equation}

Since the crest of the upper wave can have an arbitrary shift relative to the crest of the lower in general (due to an extra piece of glass near the upper slit, for example), a `phase factor' could also be added after the `+' in this expression. (This can even make the `$+$` into  `$-$', corresponding to the trough instead of the crest at a particular position). All such general expressions are called linear superpositions. The wavefunctions contain complex number quantities to accommodate the needs of the representation.  What is observable is related to the square of the wavefunction, $\left||State\rangle\right|^2$, which will involve the `interference term' consisting of the bilinear product $2|Wave\mbox{-}U\rangle\cdot |Wave\mbox{-}D\rangle^*$, agreeing with the empirical observations accurately. It turns out that one can completely forget about the picture of a physical wave, and replace it with that of the concept of a `vector' that can be added and subtracted with appropriate complex factors. In fact, modern quantum mechanics is done in the language of these `state vectors', which are equivalent to the representation with the wavefunctions.

\begin{figure}
	\centering
	\includegraphics[width=0.7\linewidth]{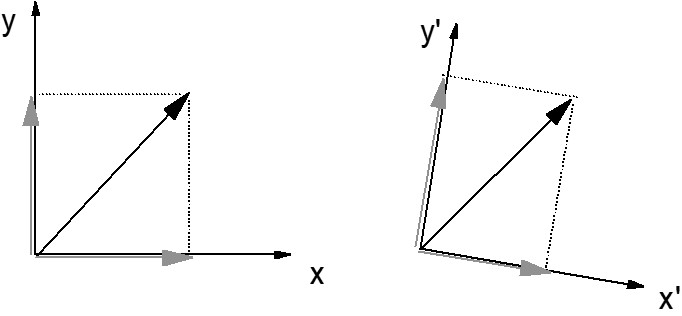}
	\caption{The concept of a vector and projections on different basis. The thick arrow is the `state vector', and the light arrows along the two directions are the projections. On the right is the same vector projected on  a different basis.}
	\label{fig:fig2}
\end{figure}

The basic idea is that the \emph{representation} of the state of a physical system is by means of a vector in an abstract space. For our purpose this vector is very similar to other familiar vectors, a directed arrow pointing in some direction in space, whose length is agreed to be one unit. (See figure \ref{fig:fig2}). 

We can choose some coordinate system in space, and  work with just two axes for clarity; x-axis and y-axis (called the `basis'), which can be any two perpendicular directions for convenience (though I refer to `space' here, the vectors of quantum mechanics are in an abstract space of any dimensions). Some immediate obvious facts about this construction can be noted: 1) the length of the vector does not depend on either rotating it, or on rotating the coordinate system, 2) the projections of the vector on the two axes depend on the rotation of the vector or the coordinate system, and the sum of the squares of the projections is, by the Pythagoras theorem, just the square of the length of the vector, 3) the vector itself can be represented as the vectorial sum of its projections, $\vec{V}=V_x\hat{x}+V_y\hat{y}$; this is a linear superposition. A general linear superposition corresponds to scaling up or down each vector and then adding them with the correct directions. 

\begin{figure}
	\centering
	\includegraphics[width=0.4\linewidth]{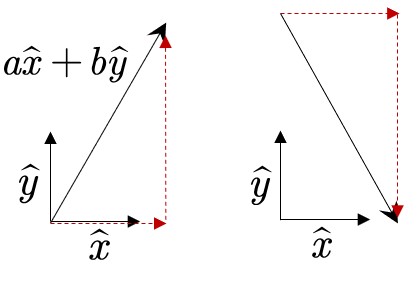}
	\caption{The meaning of a linear superposition – the vector along the diagonal is a superposition, a sum, of the other two orthogonal (perpendicular) vectors, one stretched more than the other by a scaling factor. On the right is the superposition with a negative sign for the second vector.}
	\label{fig:fig3}
\end{figure}

In classical mechanics several quantities have been described by vectors, like velocity, force etc. Given a vector quantity for a classical system we can talk about values of its projections on various directions, and we think of the physical system as possessing all these values simultaneously as its physical properties, irrespective of what measurements we make on the system as long as the measurement itself has not changed the properties. In quantum mechanics, the state vector does represent the properties of the system, but in a way very different from what one is used to in classical mechanics. The essential concept is the following. The state vector represents the physical state and the basis that we choose defines the property one wants to observe. It is sufficient to illustrate this with a simple example.

If $|s\rangle$ represents the state vector relevant for the observation of the angular momentum of a particle, then one has to choose a basis corresponding to the components of the angular momentum in various directions in space. This aspect is similar to the case in classical mechanics where if  $\vec{S}$ is the angular momentum vector for a physical system, then the angular momentum in a direction at some angle is given by the projection of the vector $\vec{S}$ along that direction. \emph{In quantum mechanics the projection of the state vector along a direction does not correspond to the value of the angular momentum in that direction}. \emph{Indeed the value is usually quantized and random in each measurement}; irrespective of what the projection is, the value measured is one of the possible discrete values, at random, say $+1$ or $-1$ in relevant units. But \emph{the square of the length of the projection can be associated with the probability of realizing a particular value of angular momentum when measured in that direction}.  This appearance of the notion of probabilities makes the quantum projections very different from the classical vector projection. In classical physics, the projection of a particular vector quantity along some direction is the component of that quantity itself, along that direction. Given the original vector, one can say with certainty that a particular value will be reproducibly obtained if projected along some direction. For observations on a quantum system, the projection is not directly related to results of individual observations. If the original state vector has unit length corresponding to a measured quantized value $s$, and if the projection has a length of 1/2, one can say that there is a probability of $(1/2)^2=1/4$, that a measurement in that direction yields the value $s$ for the projected angular momentum. So there is a shift from `value' to `probability'. 

For example, suppose the physical system (say, an electron) is prepared by some means with the value of its angular momentum along a particular direction equal to +1, in units of $\hbar/2$. It turns out that for this system, the values of the projections of the angular momentum in \emph{any direction} are always either $+1$ or $-1$, and nothing else. (There are other particles where the values could be $+1$, $0$ and $-1$). This of course is very strange, but we will not try to find a `reason' for this strange behaviour since there is no possibility of succeeding in explaining such behaviour with the familiar reasoning. Mathematically we will represent the original state vector as $|s\rangle$. Now suppose that the direction in which the system was prepared with the definite value $+1$ is along a direction that is symmetrical relative to the directions x and y, somewhat like the diagonal in figure 2 (the mathematically correct representation of this example is slightly different from the graphical example in the figure, but for our purpose the present discussion will suffice). Subsequently we want to measure the angular momentum in the directions x or y, shown as the horizontal and the vertical axes in the figure. Geometrically, the projections of the original state vector on either of these axes are equal to $1/\sqrt{2}$. Therefore, if a measurement of the angular momentum is done along the x-axis, say, then we will either get  $+1$ with a \emph{probability 1/2}, which is the square of the projection, or $-1$ with a \emph{probability 1/2}, like when tossing a coin.

Two important clarifications are necessary. The fact that one gets a particular value in a measurement, for example $+1$ along the x direction, does not mean that it had that value just before that measurement. Obviously, we had prepared the system with its angular momentum +1 along the diagonal. Yet, in a measurement along the horizontal direction, we got a value $+1$ (or $-1$). After the measurement, the angular momentum projection in the horizontal direction remains $+1$, and any further measurement in the same direction gives $+1$ with certainty. Therefore, by the Born rule, the state vector’s projection along the horizontal should now be equal to $+1$, so that its square  corresponds to the probability 1, or `certainty'. The revised description is now with a new state vector along x, with the y-component ceasing to exist -- this is called the `collapse of the state', or the `collapse of the wavefunction'. The state vector has changed suddenly on a measurement, and the vertical component has gone to zero after the measurement resulted in a definite value in the orthogonal direction. Mathematically, the situation is similar for the wavefunction corresponding to the particle passing through a screen with two slits (labelled $+1$ and $-1$). A wavefunction with the symmetrical superposition of equal components `collapses' to a single component ($+1$ and $-1$) randomly, when an observation is attempted. If we had associated the two-component wavefunction physically with the particle (as one often does with the waves of the wave-particle duality), then the vanishing of one component has the implications of the vanishing of some physical entity from space, spontaneously and instantaneously during the measurement. This is what makes the collapse of the wavefunction mysterious and troublesome.

\section{The Collapse of the Wavefunction and Other Disasters}
The `collapse of the wavefunction' is undoubtedly the most discussed and the most dreaded consequence of the mathematical machinery of quantum mechanics, as interpreted today. Valid physical states of a system are routinely described by a general sum of several distinguishable, and mutually incompatible states (if one occurs physically, the other is excluded, like the `Heads' and `Tails' states). A multi-component wavefunction with the observable possibilities $S1,S2$ and $S3$, for example, can be represented as the linear superposition 
\begin{equation}\label{key}
\left\vert \psi\right\rangle =a\left\vert S1\right\rangle +b\left\vert
S2\right\rangle+c\left\vert S3\right\rangle	
\end{equation}
A measurement on a single physical system turns up just one of the possibilities contained in the superposition, statistically, and with a probability related to the length of each vector contributing to the superposition. For instance, a measurement might result in the outcome $S2$ (with the probability $|c|^2$). But as soon as a single measurement is completed, the superposition of several states changes into just one of the states, corresponding to the value realized in the measurement. In this case, the wavefunction becomes
\begin{equation}\label{key}
	\left\vert \psi'\right\rangle =\left\vert
	S2\right\rangle	
\end{equation}
All other components vanish. If one had attributed a physical (material) reality in space and time to the components present in the superposition, then these elements had to physically vanish when a measurement is made. It is as if a physical state that was distributed and spread out into different physical states suddenly condenses into one state, when observed. In the example of the double-slit experiment, one can make an observation near one of the slits to answer the query, `through which slit is the particle passing?' If the detector registers the particle, then the particle could not be passing though the other slit. If it did not register anything, then the particle is certainly passing through the other slit. Either way, the path is determined. Having detected the path with certainty, there is no room for probabilities any more, and then the wave function or state vector component in the other path should vanish to zero immediately. Otherwise there will be inconsistency with the Born's rule which says that the probability of detection is related to the amplitude of the wave. In the case of the classical coins, just before the coin lands in the final state, its state is alternating between the two possibilities and it is no surprise or paradox for us if it finally lands randomly in one of the states. But the quantum superposition is very different – as it evolves, it is not alternating between the two states. It remains in a superposition which is neither of the two definite possibilities at any instant. Then an observation reduces it to one of the states.  In fact, there are examples where one of the states involved in the superposition is physically far away from the other state. Then a measurement annihilates one of the states even when this state is physically far away from the place where the observation is made! This is a (mild) example of `nonlocality' in quantum physics. 

\section{From Superposition to Entanglement}
Superposition is a simple concept when one thinks in terms of vectors or waves adding up on each other. Some superpositions are special in quantum mechanics, and they lead to consequences that have shaken up the very foundations of our familiar notions of causality and locality. Consider two systems, each with its own quantum states (formally called `kets') $\left|s1\right>$  and $\left|s2\right>$. Each of these could be a simple superposition of the two states $|u\rangle$ and $|d\rangle$, which are the two observable states of the system. Suppose these superpositions for each individual particle are $\left|s1\right>=\frac{1}{\sqrt{2}}(\left|u1\right>+\left|d1\right>)$  and  $\left|s2\right>=\frac{1}{\sqrt{2}}(\left|u2\right>+\left|d2\right>)$. For example, the labels might correspond to a particle spinning clockwise (u) or anticlockwise (d) when looked at from a particular direction. Since there are two particles, one can also study the `joint state' of the total system, which is written as a juxtaposition of the states, $\left|s1\right>\left|s1\right>$. The joint state would be written as 
\begin{align}
	\left|s1\right>\left|s2\right>=\left(\frac{1}{\sqrt{2}}(\left|u1\right>+\left|d1\right>)\right)\left(\frac{1}{\sqrt{2}}(\left|u2\right>+\left|d2\right>)\right)\\
	=\frac{1}{2}\left(\left|u1\right>\left|u2\right>+\left|u1\right>\left|d2\right>+\left|d1\right>\left|u2\right>+\left|d1\right>\left|d2\right>\right)
\end{align} 

In this expression the first ket $|s1\rangle$ stands for the first particle and the second one $|s2\rangle$ for the second. Each is in a state of superposition and the joint state is a simple juxtaposition of all the possibilities of the superposition. There are totally four possibilities and each one will happen with a probability $(1/2)^2=25\%$. This is a general superposition, and it is a `product state' since it can be written as a product of two factors, $\left|s1\right>$ and $\left|s1\right>$, each of which is a sum (superposition) of two factors. The important point to note is that each particle is said to have its own definite state, $\left|s1\right>$  and $\left|s2\right>$. The joint state is `separable' into $\left|s1\right>$ and $\left|s1\right>$. Now suppose we put some restriction on the possible states – say, when one particle is up the other should always be down (this might come from the requirement of some conservation principle, like that total spin is zero). This could happen due to some brief interaction between the particles. Then two of the four possibilities are prohibited, and the joint state would be 
\begin{equation}\label{key}
	\left|S=0\right>=\frac{1}{\sqrt{2}}\left(\left|u1\right>\left|d2\right>\right)+\frac{1}{\sqrt{2}}\left(\left|d1\right>\left|u2\right>\right)
\end{equation}
Now there are only two possibilities for the joint state, each with a probability of 50\%. \emph{This cannot be written as a product of the individual states $\left|s1\right>$  and $\left|s2\right>$. The joint state $\left|S=0\right>$, which is also just a superposition of two product states, is not `separable' anymore}. This is called an `entangled state' with very peculiar properties. The outcome of a measurement on each particle is entangled with that of the other; $|u1\rangle$ with $|d2\rangle$ and $|d1\rangle$ with $|u2\rangle$. \emph{Since it is not a simple juxtaposition of the two individual states, it turns out that each of the particles does not have any physical state by itself, in the quantum theory!} The independent quantum status of each particle is lost and there is only a joint state. Neither particle is in state `up' or `down', nor are they in any superposition of `up' and `down' states. However, the most general individual state possible in the quantum theory can only be some general superposition of the `u' and `d' states. Here, neither particle is in such a state. All one can say is that if one is found in `u' state, the other will be found in `d' state and vice versa. A measurement will get one of these two joint possibilities, each with a 50\% probability. If a measurement is made, the entangled state collapses into one of the two joint states,  $\left|u1\right>\left|d2\right>$ or $\left|u1\right>\left|d2\right>$. Once this happens, each particle has a definite state and the joint state is a simple juxtaposition of the individual states. The individual states and separability are regained only after a measurement on at least one particle. So, a particle that had no description in terms of a definite state before the measurement acquires a definite state as a result of measurement. Moreover, only a measurement on one of the particles is sufficient – the other particle acquires a definite state automatically without any measurement or observation on it! For example, if the measurement shows that the first particle is in the state $\left|d\right>$, then definitely the joint state is $\left|d1\right>\left|u2\right>$, and not $\left|u1\right>\left|d2\right>$. This means that \emph{the second particle, on which no observation was made, is now certainly in the state $\left|u\right>$ after the measurement on the first particle, though neither had any specific state whatsoever before the measurement on the first particle}. 

Therefore, the quantum entanglement in the state makes the properties of the particles inseparable, and one gets definite properties that are emergent only after an observation. 
\emph{The bizarre consequence is when the two particles are not in the same spatial region}. The particles can be separated after the interaction that made the entangled state. Then a measurement on one particle instantaneously gives it a definite state \emph{and also a definite state to the other particle that is far away from the first particle}! This is the much discussed violation of locality in quantum mechanics. On the other hand, if we assume that one measurement should not influence the physical state of the other particle that is spatially separated, then we see that quantum mechanics has no description of such independent states. Quantum mechanics corresponds to an incomplete description because its states can be affected by an arbitrarily distant measurement, whereas the requirement of locality is that a physical state in what ever form should not be  influenced instantaneously by processes that are spatially distinct and separated. It was Einstein who pointed this severe difficulty in the well known Einstein-Podolsky-Rosen (EPR) paper in 1935 \citep{EPR1935}. Schr\"odinger's papers \citep{Schrodinger1935} on the properties of such joint states, entanglement, and its crucial place in the problem of measurement were partly inspired by the EPR paper.

\section{The Schr\"odinger Cat}
We now come back to the description of the Schr\"odinger cat. In his paper~\citep{SCat-Eng} that summarized the ``Present state of quantum mechanics'' he wrote, 
\begin{quote}
	One can even set up quite ridiculous  cases... The wavefunction of the entire system would express this by having in it the living and the dead cat (pardon my expression) mixed or smeared out in equal parts.
\end{quote} 
The colloquial expression ``the living and the dead cat mixed or smeared out in equal parts'' is the source of a serious confusion to those who have not seen Schr\"odinger's precise discussion in a follow up paper \citep{Schrodinger1935}.

There are descriptions of the general physical situation implied here without any mention of a live body, and yet completely equivalent to it from the point of view of the mathematical theory of quantum mechanics. I mention one such by Einstein, written in 1948 \citep{Schilpp}:
\begin{quote}
	As far as I know, it was E. Schr\"odinger who first called 
	attention... Rather than 
	considering a system which comprises only a radioactive atom 
	(and its process of transformation), one considers a system 
	which includes also the means for ascertaining the radioactive 
	transformation -- for example, a Geiger-counter with automatic 
	registration-mechanism. Let this latter include a registration-strip, moved by a clockwork, upon which a mark is made by 
	tripping the counter. True, from the point of view of quantum 
	mechanics this total system is very complex and its  
	configuration space is of very high dimension. But there is in principle 
	no objection to treating this entire system from the standpoint 
	of quantum mechanics ...
	The reason for the introduction of the 
	system supplemented by the registration-mechanism lies in the 
	following. The location of the mark on the registration-strip 
	is a fact which belongs entirely within the sphere of macroscopic concepts, in contradistinction to the instant of disintegration of 
	a single atom. 
\end{quote}
The way Einstein describes the situation stresses the mathematical equivalence of the example of the Schr\"odinger cat, and the quantum measurement problem that involves a macroscopic `apparatus'. It is possible to represent systems like the Schr\"odinger cat mathematically in a straightforward way; the initial state of the system while the cat is being placed in the box along with the undecayed atoms is   $\left|S_i\right>=\left|A+\right>\left|CL\right>$ where we denoted `$A+$' for the undecayed atom, and `CL' for the live cat. This is  a product state. We have written the total state as simply the state of the atom AND the state of the cat together. If there is definite knowledge that the atom has decayed, then of course the cat is dead soon. So, given the knowledge that the atom has decayed, the joint state is  $\left|S_f\right>=\left|A-\right>\left|CD\right>$, where `$A-$' stands for the decayed atom. The peculiarity of quantum physics is that if there is no definite knowledge of the state, when the box is closed with the cat, atoms and the poison inside,  the state of the system can be some (time dependent) linear superposition of these two possible joint states,
\begin{equation}\label{key}
	\left|S_t\right>=a(t)\left|A+\right>\left|CL\right>+b(t)\left|A-\right>\left|CD\right>
\end{equation}
Now, what does this mean? Unlike the other two states ($\left|S_i\right>$ and $\left|S_f\right>$), one cannot say whether the atom has decayed or not (or equivalently, whether there is a mark on a paper strip after the Geiger counter). Correspondingly, one cannot say whether the cat is dead or not. It will be incorrect to say (in the quantum theory) that the atom is either decayed or not and the cat is either alive or dead. That is not what the state says. Most importantly, the atom is not in a superposition of its two possible states, and the cat is not in any superposition of the states `alive' and `dead'! But, there are no other individual states in the quantum mechanical description -- we have exhausted all possibilities. The joint state clearly says that the state is a superposition of the two joint possibilities, and that neither the atom nor the cat is in any definite state. \emph{In particular, it does not say that the atom is both decayed and not, or that the cat is both dead and live}. (It is this incorrect description, \emph{in which the quantum cat is in a superposed state of both alive and dead}, that is often debated as the `paradox of the Schr\"odinger cat'!)

We are talking about the state or the wavefunction representing the state of the system in some abstract space. The only definite thing quantum mechanics says at this point is that IF a measurement is performed (opening the box), one will either see the state $\left|A+\right>\left|CL\right>$ (the atom undecayed AND the cat alive), OR the state $\left|A-\right>\left|CD\right>$ (the atom decayed AND the cat dead), corresponding to the definite states $\left|S_i\right>$  and  $\left|S_f\right>$. It is very important to note that the neither the atom nor the cat are in states of superposition separately. \emph{The cat is  definitely not in the state $\alpha \left|CL\right>+\beta \left|CD\right>$. The cat has no state at all}! So hasn’t the atom. But together they have some quantum state. And that state has no implication on what the state of the cat or the atom is until a measurement is made. Therefore, typical statements, commonly made by almost everybody, like `the cat is both dead and alive’, or the `cat is in a superposition of being dead and alive’ etc. are wrong. They completely miss the meaning of the simple entangled state in quantum mechanics. It is unfortunate that such serious misrepresentation of a fundamental point is made in spite of Schr\"odinger's clear description \citep{Schrodinger1935} (italics added), 
\begin{quote}
	When two systems, of which we know the states by their respective representatives,
	enter into temporary physical interaction due to known forces between
	them, and when after a time of mutual influence the systems separate again, \emph{then
		they can no longer be described in the same way as before, viz. by endowing each
		of them with a representative of its own}. I would not call that one but rather the
	characteristic trait of quantum mechanics, the one that enforces its entire
	departure from classical lines of thought. By the interaction the two representatives
	(or $\psi$-functions) have become entangled.
\end{quote}

\section{Common Misconceptions}
I would like to summarize the common misconceptions about the Schr\"odinger’s cat parable before proceeding further. Even professional physicists often tend to loosely say, either because they assume that the meaning is understood or because some of them genuinely lack clarity, that the superposed state means the cat is both dead and alive. I quote from some recent technical papers: 

``...an unfortunate cat is placed in a quantum superposition of being dead and alive… This situation defies our sense of reality because we only observe live  or dead cats, and we expect that cats are either alive or dead independent of our observation''.

``Applying quantum theory to our macroscopic world, Schr\"odinger proposed a thought experiment with an unfortunate cat lingering in the twilight zone between life and death…'' 

``The paradox arises because the atom, being a microscopic object, must be described by quantum mechanics. After one hour, and before it is observed, the atom is in an equal superposition of being decayed and undecayed. However, if quantum mechanics is a universal and complete theory, it must describe the whole system. And, since the state of the cat is correlated with the state of the atom, the cat must also be in a superposition of being dead and alive. This clearly contradicts our everyday experience of cats! This sort of apparent contradiction arises whenever the state of macroscopic objects is correlated with that of microscopic objects. It comes about because in our experience of our macroscopic surroundings, objects are either in one state or another, but never in a superposition of several states at the same time.''

``…the quantum state must ultimately involve such a complex number superposition of a dead cat and a live cat: so the cat is both dead and alive at the same time!'' …``Of course such a situation is an absurdity for the behaviour of a cat sized object in the actual physical world as we experience it''. 

Any statement which asserts that the cat is both dead and alive at the same time in the entangled state (or even in a simple superposition) is the misconception that is shared by a large number of people. The confusion is at two levels:  1) the central point about entanglement that neither system has an individual quantum mechanical state whatsoever is missed, 2) even the notion of superposition in standard quantum mechanics or in quantum mechanical experiments known so far never implied that the physical entity itself carried both the properties represented by the state at the same time. This needs to be clarified once and for all.

In standard quantum mechanics, if an electron has its spin pointing in the up direction, it can be described also as an equal superposition of states corresponding to `right' and `left' in any horizontal direction. If a measurement is made on the electron with an apparatus sensitive to the spin direction in the horizontal direction, then one gets `left' or `right' with equal chance, though originally it was prepared in the state `up'. In fact, after getting `right' or `left', if an attempt is made to measure it in the vertical direction, one ends up getting `up' or `down' with equal chance! In any case, one does not say that the particle with spin `up' is both in the `left' and `right’' direction at the same time. It is in the `up' state. And \emph{nobody has any empirical experience of seeing an electron with its spin both pointing `left' and right', in spite of its microscopic status!} What is seen empirically is that an ensemble of electrons contains both `left' and 'right' pointing electrons. This example is mathematically and physically similar to that of any superposition in quantum mechanics.

The relevant question then is why isn't there any physical state in the macroscopic world that is a \emph{superposition of states} corresponding to being alive and being dead. That would be some new state of `being' that has equal projection to the state representing `being dead'  and `being live'. But that still does not mean that some organism in that state of being is both dead and alive.  
Indeed, nobody has seen an atom both decayed and undecayed. Nobody has seen an electron at two different places at the same time. The superposition one is always referring to is the superposition of the wavefunctions, or the state vectors, representing the two distinguishable states of the quantum entity. The connection between the quantum entity (whether it is an atom or a cat) and the wavefunctions representing the various states of being is not a physical identity or equivalence. When an observation is made ``to see', only one of the observable possibilities are seen, with the probabilities corresponding to the magnitude of the wavefunction for each state making up the total wavefunction. Theoretically it may be correct to say that the cat cannot be described or represented as either dead or alive, but it is certainly incorrect to say that the cat is neither dead nor alive, or to say that the cat is both dead and alive. The total wavefunction in superposition is often interpreted as meaning both the above statements, and clearly, from a logical point of view, this lack of uniqueness goes against such interpretations. 

Pressing a bit further, one might want to ask the question during some point of time in the experiment, ``is the cat alive or dead inside the box?'' Quantum mechanics has no way of answering this question. \emph{In fact QM never gives a definite answer to a question that is asked about the exact physical state of a single physical entity}. It just gives the probabilities of the cat being found dead or alive when the box is opened. \emph{In fact, quantum mechanics has nothing to say about any individual cat}. The predictions of QM are statistical, and if the experiment is performed repeatedly, the fraction of cats that are dead and the fraction that are alive will match well with the squares of the corresponding wavefunctions. 

So, Schr\"odinger was not discussing a paradox of quantum physics. He was just discussing a ridiculous example that needed the correct interpretation. If one associated the wavefunction, as a physical equivalence, to the quantum entity, then one has to say that the quantum entity itself is in a superposition and therefore could also be in a superposition of being dead and alive at the same time. Therefore, this ridiculous example warns us from wrongly interpreting the wavefunction. As Schr\"odinger himself said, ``It is typical of these cases that an indeterminacy originally restricted to the atomic domain becomes transformed into macroscopic indeterminacy, which can then be \textit{resolved} by direct observation. That prevents us from so naively accepting as valid a ``blurred model'' for representing reality.'' One need not bring up the example of the cat for being sufficiently warned. When an atom passes through a screen with two closely spaced slits, the quantum wavefunction after the slits is very similar to the superposed wavefunction we can write for any macroscopic object. But this does not mean that the atom itself got split at the slits, or that there are atoms in both paths after the slits – that would violate a whole lot of known physics, including the fundamental conservation laws. So, while we do not know what is split into two at the slits (it is called the wavefunction!) what we know for sure is that the atom itself is not split. Same logic applies to the Schr\"odinger’s cat.

\section{Why is the Cat Special?}
Schr\"odinger chose to describe the issues that involve a linear superposition of states representing life and death. This was perhaps the burlesque aspect he refereed to, because it is otherwise common and natural in quantum mechanics to have exactly this kind of entangled superpositions whenever two systems interact.

This makes the original Schr\"odinger’s cat very different from the  `cat states' described in most of the current discussions, though by and large physicists have ignored this difference. The cat had two special aspects- it is really big on the scale of atoms, and it is a living being. We also think it has a reasonably developed intelligent brain, thought and awareness.  Each of these aspects make the original Schr\"odinger cat state different from the states experimentalists deal with while discussing the Schr\"odinger cat problem. It is a routine context in experimental and theoretical quantum mechanics to come across entangled states. Consider an atom entering a small cavity in which there is no light. The atom enters in an excited state ($A+$). While passing through the cavity the atom has some probability to decay ($A-$) and deposit a photon in the cavity ($C+$). It also has a similar probability not to decay during the passage ($C-$). Now this situation is very similar to the case of the Schr\"odinger’s cat, except that we are talking about a microscopic change in the state of the macroscopic cavity (one photon deposited or not). The corresponding mathematical states are $\left|A+\right>\left|C-\right>$ and $\left|A-\right>\left|C+\right>$. The entangled state is simply the superposition of these two joint states. Again, the atom or the cavity does not have a specific state, and neither of them is in a superposition of any specific states. But the two together are in a superposition of possibilities of togetherness. One may use this inanimate, soul-less, consciousness-less state as a comparison to the original Schr\"odinger’s cat state. First of all, though the cavity itself might be macroscopic, there is only a microscopic change between its two possible states, differing by one photon. \emph{The live and dead states of an animal differ in so many aspects that it cannot be considered as a microscopic change}. Then, there is the aspect of the irreversible relation between life and death. While it is natural and relatively easy to evolve from a state of being live to a state of being dead, the reverse evolution is never seen. So, in the preparation of the Schr\"odinger cat state, one cannot start with a dead cat and then hope to get a superposition of states corresponding to being alive and dead. But in the case of atom and the cavity, one can start with a cavity with one photon and an atom in the ground state (the second state above) and can have a superposition with a state in which the cavity has zero photon and the atom is in the excited state. This essential irreversibility is unique to the Schr\"odinger’s cat state. On the other hand, I should state that technically such irreversibility can be mimicked by using a bad cavity where the photon cannot be held for sufficiently long time. 

The third difference is perhaps the most significant from the point of view of the interpretation of quantum mechanics. Whether or not the observer outside the closed box has decided to make the observation, the conscious animal inside will `know' when it is dying, and when the glass vial with poison has broken. As far as the animal is concerned, though totally unaware of Hilbert spaces and quantum superpositions, or even of Schr\"odinger, its own state or the state of the glass vial containing the poison is known with certainty. If one insists on an abstract representation, the state is either one or the other, and never a superposition, just as for an observer sitting inside the box and watching carefully, there is no quantum superposition. One may propose that for the observer outside who is forced to use the quantum formalism, nothing has changed if a secret observer or the cat himself is making the observations inside the box – neither the mathematical representation nor the physical results in measurements could change, whereas the description of the system is completely different for the observer inside and for the observer outside. This then would show that the wavefunction is not a property related to the state of the physical system – it merely represents the unobserved possibilities and \emph{subjective information} in a way consistent with various statistical probabilities inherent in the system. But this view turns out to be incorrect. If there is any means, hidden or known, of breaking the superposition – be it a secret observer inside the box, be it the consciousness of the cat, or be it a purely inanimate surveillance system that registers the state of the cat – then the superposition is broken and the system is reduced to a definite state. As soon as the `information' is accessed physically from the quantum system once, the wavefunction changes and no subjective element is left in the description. We will now discuss this important aspect in some more detail.

\section{The Schr\"odinger Cat and the Quantum Measurement Problem}
The observation of a physical quantity in a system involves a subtle conceptual issue, which becomes a core physical issue of quantum mechanics itself, when fully formulated. Every observation or `measurement' needs an `apparatus', which we usually designate as the experimental set up. This is invariably a macroscopic physical system, but could also include the live `observer'. In any case, let us focus first on the microscopic quantum system and the measuring apparatus. \emph{The process of the measurement involves a physical interaction between the system and the apparatus}. We take for granted that the apparatus is a `classical object', playing its role of extracting the information on the observable quantity, without entering in the theoretical machinery of quantum mechanics. But such an assumption amounts to admitting that quantum mechanics is not a universal theory, and that it is applicable only to microscopic system. Then, one is asserting that a separate and distinct theory describes macroscopic objects like the apparatus. On the other hand, if we claim that quantum mechanics is the universal theory applicable to the dynamics of all objects, then there are serious consequences for which no solution has been found. This is the quantum measurement problem.

It was Schr\"odinger who  first discussed the core issue. In the same paper that discussed the concept of entanglement, he wrote, 
\begin{quote}
	By the interaction the two representatives
	(or $\psi$-functions) have become entangled. To disentangle them we must
	gather further information by experiment, although we knew as much as anybody
	could possibly know about all that happened. Of either system, taken
	separately, all previous knowledge may be entirely lost, leaving us but one
	privilege: to restrict the experiments to one only of the two systems. After reestablishing
	one representative by observation, the other one can be inferred
	simultaneously. In what follows the whole of this procedure will be called the
	\emph{disentanglement}. Its sinister importance is due to its being involved in every
	measuring process and therefore forming the basis of the quantum theory of
	measurement, threatening us thereby with at least a \emph{regressus in infinitum}, since
	it will be noticed that the procedure itself involves measurement.
\end{quote}

Thus, every measurement involves the interaction of the system and an apparatus, each of which will be described by its own $\psi$-function, before the interaction. This will be some general superposition of the possible physical states, in each case. However, the interaction results in the entangled state, exactly of the same nature as the joint state of the atom and the cat in the example of the Schr\"odinger cat. This will be an entangled state of the quantum system `s' and an apparatus `A', 
\begin{equation}\label{key}
	\left|\psi\right>=a\left|s+\right>\left|A+\right>+b\left|s-\right>\left|A-\right>
\end{equation} 
Then, neither the system nor the apparatus has any definite physical states of its own! There is only a joint state, with less information, as Schr\"odinger noticed. The problem is even more severe. To determine the state of the apparatus, which is the measurement, the observer `O' has to access the `reading', which involves an interaction of the observer with the apparatus. However, if the theory of quantum mechanics applies to everything including the observer, all that we will manage is the larger entanglement involving the system, the apparatus, and the observer! 
\begin{equation}\label{key}
	\left|\psi'\right>=a\left|s+\right>\left|A+\right>\left|O+\right>+b\left|s-\right>\left|A-\right>\left|O-\right>
\end{equation}
Adding more observers simply enlarges the entanglement. This is the \emph{regressus in infinitum} Schr\"odinger referred to. There is no way to realize a final observation of a definite disentangled physical state, if quantum mechanics is the correct theory and if it is a universal theory. This is a devastating conclusion, and that is why the quantum measurement problem is treated as the foremost unsolved problem in the formulation of quantum mechanics.
\section{Wigner's Friend}
The act of observation on a quantum particle that is described by a superposition of states reduces the superposition to just one of the terms in the superposition (in general the new state could be a different superposition, but here we will consider the case when there are only two possibilities and the measurement is designed to reduce the superposition to one of the states). When that happens, the system is described by a definite well defined state. This reduction from the general superposition to one of the component states after a measurement is called reduction of the wavefunction or more graphically, the collapse of the wavefunction. In the context of the Schr\"odinger’s cat, opening the box and looking into is the act of measurement and this collapses the wavefunction to one of the two possibilities. Let us say in a particular experiment the reduction took place to the state $\left|CD\right>$. Then the atom has definitely decayed and the cat is dead. But QM does not allow one to conclude that this was the state of the system even the smallest time interval before the box was opened. Before looking into the box,  the system is described by the superposition of joint states and after opening it, the system will be described by one of the joint states. The interesting thing about the interpretation of QM is that if there is no superposition, one could claim without inconsistency that the system is in that particular definite physical state, and in this special case one can attribute a physical equivalence between the wavefunction (state) and the system without inconsistency. 

In 1956, Eugene Wigner modified the Schr\"odinger parable to include more characters \citep{Wigner-friend}. Instead of an observer directly looking in, suppose a `friend' is asked to keep a watch and then answer queries about the outcome in the box. Again one can think of the friend as a part of the system and formally write a wavefunction to include the friend. The outcome observed by the friend is correlated with some state of neurons in his brain, and one can imagine that the total wavefunction now is the tri-partite entangled state,
\begin{equation}\label{key}
	\left|S_t\right>=a(t)\left|A+\right>\left|F+\right>\left|CL\right>+b(t)\left|A-\right>\left|F-\right>\left|CD\right>
\end{equation}
For the query by the observer, the friend answers Yes or No ($+$ or $-$) and then the system collapses to one of the definite states. 

The complication arises because the friend is supposed to have a conscious mind, and his observing the cat has the same role physically as the final observer looking at the cat. Therefore, if the wavefunction collapse is a physical process then it would have happened when the friend looked at the system, and the wavefunction written above is not valid. The friend is never to be described in a superposed state. His consciousness has completed the act of observation and his own ideas about the description of the system and himself are to be described even within quantum mechanics by definite states that are not in superposition. But as far as the outside observer is concerned, consistency demands ignoring the consciousness and writing the formal wavefunction; this gives results consistent with a limited set of experiments that asks only the probabilities regarding the two states `dead' and `alive'. 

The cat itself may have a conscious mind, being a living and highly evolved animal. Then writing a superposed wavefunction to describe a cat will be absurd. QM is supposed to be applicable to the entire world, irrespective of the size and weight of the objects, but examples like that of Wigner’s friend prompt us to think that there is perhaps a very important divide between macroscopic physics and microscopic physics. Perhaps there is a grey area as well, a transition region, where quantum mechanical description slowly merges into the classical physics description of particles and cats where superposition has no meaning. 

There is a discussion of these issues by B. d’Espagnat, who explored the relation between physical reality  and `mentality' \citep{Espagnat2003}. Essentially, he finds the possibility of relating the hidden variables in the de Broglie-Bohm version of the quantum theory to an `internal state of consciousness', thereby bringing the consciousness of Wigner’s friend and the unobserved paths of the particles on a similar level of physical discussion. He makes these internal states of consciousness elements of physical reality, but distinguishes them from elements of empirical reality by stating that the internal state of consciousness is not communicable. These two notions of consciousness are then linked through mechanisms of physical interactions. A closer look at such approaches reveals that no new significant insight is achieved, and that in effect a different set of words is used in a description that remains conceptually and empirically similar to the older discussions.

From the earlier discussion of entanglement, and the \emph{regressus in infinitum} that Schr\"odinger mentioned, it is clear that Wigner's introduction of a conscious being in the process of observation was rather naive. It amounts to the statement (by a physicist) that there are  elements in physical nature that cannot be described within physical theories like quantum mechanics.

\section{Is Quantum Superposition Subjective and Information Based?}
The conscious cat is never in a superposition. But for the unaware observer the superposition is broken only after the observation. Consciousness enters the formalism, though there is no argument that can prove that it enters the physical process. The question arises whether the quantum formalism can survive up to conscious observers capable of first person self-observations of their own state. This has been already addressed in the example of Wigner's friend. An examination of this issue in detail in not intended in this article. However, it is relevant to describe the essential ideas concisely.

The wavefunction of a physical system is intimately related to the physical configuration of the experimental apparatus that observes the physical system. For example, for a light beam that approaches a screen with slits, the subsequent superposition of states corresponding to the passage through each of the slits crucially depends on the properties of the slits. Therefore, to describe the quantum state of the particle, the conscious observer should know the experimental system well; otherwise the quantum state that he writes to describe the physical processes and observations could be incorrect and then the observations will not agree with the theoretical description. Now let us consider the situation where a second observer puts a detector near one of the slits, but does not detect a particle. Therefore, the particle has gone through the other slit, and for this observer the state of the particle at a distant screen is not a superposition of the states corresponding to the passage through each of the slits. As far he is concerned, the wavefunction corresponds to the passage through only one of the slits, and then there is no superposition. He concludes also that there will not be any interference at the screen. It is possible to do this kind of a measurement without destroying (absorbing) the particle itself, and therefore the observer who is unaware of this prior observation will not be able to say whether somebody is observing the particles or not by merely looking at the number of particles that reach the screen. Also, if the second observer makes observations to see through which path each particle arrives, he will get the 50-50 probability expected from the experimental situation, and as predicted by quantum mechanics itself. Transferred to the case of the cat, an observer sitting inside the box can see at each moment whether the cat is dead or alive, whereas the person outside cannot. Irrespective of this situation, the quantum mechanical prediction regarding the probability of observing it dead or alive comes out correctly, as calculated from the superposed state. At this point one might think that it is reasonable to associate the superposition with `information' or `knowledge'. The observer inside the box has the information that the cat is either dead or alive, and therefore for him its description is never in a superposition. But for the observer outside, the information on the state of the cat is not available till an observation is made, and therefore he could write a superposition and predict from that the probability for seeing the cat alive or dead. If this were the case, then we could conclude that the wavefunction represents information as obtainable by a conscious being, and this might resolve a number of issues in the interpretation of quantum mechanics. Indeed, there are several new generation quantum physicists who think along these lines, especially those who work in areas related to quantum information and quantum computation. But this is an incorrect view. In the case of the double-slit experiment, the first observation that determined the path of each of the particle destroys the entire possibility of interference at the screen, irrespective of whether the observer who set up the screen to do the experiment has the knowledge of the state of the system. \emph{If he writes a superposition for the state because of the lack of his knowledge of the first observation, he will find a contradiction with his quantum mechanical prediction.} No matter what he writes for the state, there will not be any physical interference pattern. That possibility is permanently destroyed after the first attempt at the determination of the path of the particles. In fact, the interference possibility is destroyed even if the first observer never looked at the information collected near the slits. The very positioning of the working detectors capable of detecting the path destroys all subsequent possibility of interference and hence the possibility of a coherent superposition is destroyed right after the first detectors. This example clearly shows that a coherent superposition has nothing to do with the information in the brain of a conscious observer.

For the case of the cat, therefore, if there were a possibility to write a superposition state corresponding to its physical state, any sort of experimental apparatus that can interact with the cat and determine with certainty its state of being will destroy the coherent superposition. However this does not alter the probability of observing it alive, just as the probability of detecting the particle near one of the slits remains 50\% even for the second observer. Hence we conclude that observing it alive or dead is not the signature of a coherent quantum superposition. It is the possibility of an interference. This is related to the possibility of an observation of a state of being that is different from being alive and being dead. In a sense, every living being has the presence of death, among other things, inherent in life, and this is indeed the natural state. But we are discussing about formalising a quantum physical state that is described as superposition of the states representing life and death. Since it is impossible to elaborate on this further I will stop this part of the discussion here. 

\section{Macroscopic Superpositions}
At this point a second kind of the Schr\"odinger cat should be mentioned, which is different from the original Schr\"odinger cat, but quantum theoretically related to it. In fact in many experimental situations physicists attempt to create this second kind of `cat'. It is our general notion that microscopic objects like atoms and electrons can be such that their physical states could be in a linear superposition. For example, the electron has a spin angular momentum, and the two possible states are the spin pointing up and the spin pointing down. But a superposition of these two states is also a valid state. The question naturally arises whether it is possible to have a linear superposition of the states of a macroscopic object: a cricket ball can be spinning in such a way that it is either in a state of `off-spin' or in a state of `leg-spin' – is it possible to have a superposition of these two states? Such a possibility will imply that the ball will not be spinning in either of these states till an attempt is made to decide which direction it is spinning – and the attempt will reveal with equal probability that it does the leg-spin or the off-spin. While spin bowlers might love to learn this quantum spin superposition, this does not seem possible. All our experience shows that macroscopic objects always are in definite states whether or not we are observing them. This is of course a belief, but there are reasons for holding such a belief. However, quantum mechanics, as a formalism, allows with complete generality a linear superposition of states, irrespective of the size of the object.  In principle, the state of a cricket ball could be in a superposition of the leg-spin and the off-spin. The two important questions to be answered are a) is it possible to prepare (and observe) such states, b) if not, why is it impossible?

Quantum physicists often call a superposition of states corresponding to a macroscopic object, like a ball, or even more modestly, a pearl, a Schr\"odinger cat state. There is no entanglement, the system is in some definite state, but it is also in a superposition of states when considered in some specific basis. An example could be a current flowing in a loop of wire, and the possible states are a clockwise current and an anti-clockwise current. A Schr\"odinger cat state in this situation will be a state corresponding to a superposition of states corresponding to a clockwise and an anti-clockwise currents (again, this does not mean that the currents are actually flowing clockwise and anticlockwise).
Going one step further, one can ask the question whether the cat or any living being could be in a state of superposition of states corresponding to being alive and being dead. In other words, is the state $\left|\psi\right>=a\left|CL\right>+b\left|CD\right> $ possible?
\begin{figure}
	\centering
	\includegraphics[width=0.3\linewidth]{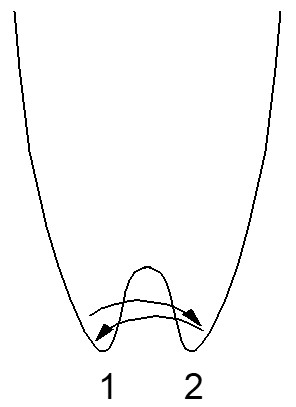}
	\caption{A two-state system with the possibility of quantum tunnelling between the states 1 and 2. If the system is initially prepared in the state 1, it has some probability to leak into the state 2, and the probability can oscillate back and forth. The quantum description of the state of the system is in terms of a superposition. Classically, if the system is prepared in the `well' of state 1, it remains there for ever.}
	\label{fig:fig4}
\end{figure} 

If a macroscopic familiar physical object has two possible distinguishable states labelled as 1 and 2 (Heads and Tails for a coin, for example) we will first assume that there is a quantum mechanical description of these states employing state vectors, and we call them $\left|1\right>$ and $\left|2\right>$. So far the correspondence is formal. But once the applicability of quantum mechanics is assumed for objects like coins, the state $a\left|1\right>+b\left|2\right>$ becomes legitimate. Classically this has no operational meaning. In quantum mechanics, this implies that when observations are made in repeated experiments the distinguishable states 1 and 2 will show up randomly with the probabilities $a^2$  and $b^2$. Only when $a=b$,  the values will show up with equal probabilities of 1/2 each, as in the case of a normal coin. The quantum state implies an even more bizarre consequence; it implies that the coin or the two-state physical object can be prepared in special situations such that if we start with the state $\left|1\right>$, there is a nonzero probability that after some time it is in state $\left|2\right>$. This is called tunnelling. Indeed, the radioactive atom naturally is in a situation where if we start with the undecayed atom, its quantum state becomes a superposition of the states corresponding to decayed and undecayed atoms. The same physical picture should apply to (special) coins as well if they obey quantum mechanics. In the case of the atom, once the decay happens, the chance that the decay products will combine back to form the atom again is essentially zero. This is because the decay products are allowed to escape. One can imagine situations where the radioactive atom as well as the decay products are confined in space such that there is some chance that the atom is restored to its original state. When that is possible the behaviour of the system becomes oscillatory, in the sense that the decay is followed by a restoration of the atom and then again a decay, and so on.

Therefore, \emph{one experimental signature of being able to prepare the superposition of macroscopic states is the oscillations between two possible states} as a result of what is called the `macroscopic quantum coherence'. This is similar to certain phenomena involving superconductivity, like the Josephson effect. But there, what oscillates is the density of a large number of paired electrons. For a single particle, the quantity that oscillates is the probability. Can the quantum mechanical probability oscillate between two distinct states for a macroscopic system? In the case of the coins this will mean that the certainty of seeing the coin `Heads' will slowly change to the certainty of seeing it `Tails' and then back to Heads and so on, provided the state of the coin can tunnel between these two states somehow. Most of the experiments seeking to study the Schr\"odinger’s cat problem look at this kind of physical effects in macroscopic quantities, like a current.

There are two views that one can take at this point. One is to say that quantum mechanics applies to all objects, and that coins can certainly tunnel between two states. But since the coin is big on the atomic scale, its quantum mechanical parameters like the de Broglie wavelength are extremely small compared to its own size as well as to the size of all scales in the problem. Therefore, the probability of tunnelling is extremely small – so small that there is practically no chance that one will see this tunnelling over any reasonable time scales. Indeed a calculation using standard quantum rules shows that this explanation is certainly plausible. Then the problem is closed.
Another view is that coins are fundamentally different from atoms, and they obey different rules of dynamics. There is a transition of rules from the microscopic world to the macroscopic world. The research on this possibility is considered very important, but the progress is slow.

\section{Experiments on the Schr\"odinger Cat}
Now I will briefly review the experimental activity related to the discussions on the Schr\"odinger’s cat.  A much discussed early experiment that dealt with a superposition at a scale larger than the mass scale of electrons or atoms was a double-slit experiment that was performed using the molecule Carbon-60 or the `Bucky-ball' \citep{Arndt1999}. Each molecule contains 60 Carbon atoms and hence one can claim that its scale is much larger than what one is used to in quantum mechanical or optical double-slit experiments. An interference pattern was obtained in this experiment, as expected if quantum mechanics applied to molecules of this size. Of course, nobody was surprised at this result, and it is treated as an important step towards testing for superpositions on an even larger scale as in the case of double-slit interference experiments with molecules of DNA, or even `live' systems like a virus. These states are called `mesoscopic' Schr\"odinger’s cat states. Physicists retain the names despite scaling down enormously the physical system as well as the content of the concern expressed by Schr\"odinger. It should be obvious that a genuine Schr\"odinger cat state is an entangled state with another system \citep{Schrodinger1935}, and a double-slit experiment with a simple superposition, $a|\psi_1\rangle +b|\psi_2\rangle$, is not an experiment on a Schr\"odinger cat state.

A notional analogue of the original entangled state that Schr\"odinger himself discussed has been created at the atomic scale, as a superposition of two states corresponding to two widely separated spatial positions of a single Beryllium ion in an electromagnetic trap. The ion oscillates in the trap by a distance of the order of 100 nanometres or so, large compared to the typical atomic size or the size of the quantum uncertainty in each of the position states \citep{Monroe1996}. If a classical particle oscillates in such a trap, it will have some definite position at each instant, and the extreme positions in the trap are accessed periodically back and forth. When the particle is at one extreme it is not at the other. If we denote these two extreme positions as $x1$ and $x2$, and two other internal states of the ion as (up) and (down), the experimenters managed to put the ion into a state of superposition of states corresponding to the position $x1$ entangled with the (up) internal state of the ion and position $x2$ with the (down) state of the ion. Of course when a measurement is done, the ion would be found at only one of these positions, with the corresponding internal state. However, this is the entanglement of quantum states in a single quantum system (single ion), unlike the entanglement of states of two physically distinct systems (the atom and the cat) in Schr\"odinger's example. Mathematically, the state is identical to the Schr\"odinger’s cat state, because it is an entangled state, though there is nothing `ridiculous', or even macroscopic, about the one produced in the laboratory.

There is another class of experiments in which coherent oscillations between two possible and classically distinguishable states of millions of electrons in a superconducting device are monitored. The superposition corresponds to that of a clockwise and an anti-clockwise current of large number of electrons. Classically it is impossible to think of a current flowing in both directions at the same time in the same wire. Quantum mechanically, the state of the electrons can be a superposition of states corresponding to currents in opposite directions. There is no entanglement involving a large number of particles here. The experimental signature in these situations is the existence of a step in the energy levels of the system and this has been measured \citep{Friedman2000,Wal2000}. Again, these experiments test for the applicability of quantum mechanical superposition principle when the physical system has a size (number of particles) much larger than what one is used to in a typical quantum mechanical experiment. Except for this size factor, the system is similar to other quantum systems like atoms or electrons and does not contain the subtleties of the original discussion by Schr\"odinger. Further, the \emph{state of the original Schr\"odinger cat consists of one atom and one cat, and not a coherent ensemble of identically prepared atoms and cats}. There is a vast conceptual difference between the two situations.

\section{Irreversibility of Death and the Schr\"odinger Cat}
The examples of superpositions discussed above contained the possibility that due to quantum tunnelling, if the state is prepared in one of the states, it has a probability to evolve into the other state; eventually the system oscillates back and forth between the two possible states. As we discussed, \emph{what oscillates back and forth regularly is the probability, and not the physical system itself}. A physical system with its state represented by the state vector in a superposition can only make quantum jumps between the two possible states, randomly. But, if a large number of similar physical systems are prepared in one of the states, after some time more and more physical systems start appearing in the other state and the number in the other state increases smoothly. Then the process reverses and keeps on oscillating. But this is possible only if the states are reversible – there should not be any fundamental limitation in reaching from one state to the other. While it is physically possible and indeed easy to go from a state of being alive to a state of being dead, the other possibility does not exist. Therefore, the superposed state of the Schr\"odinger’s cat differs subtly, in a fundamental sense, from the kind of superposed states people create in the laboratory. 

\section{The Quantum Zeno Effect: The Eternal Cat}
The quantum Zeno effect  was highlighted by E. C. G. Sudarshan and B. Misra \citep{Misra1977}. The original Zeno (of Elea) paradox referred to the apparent impossibility of motion, and there is not a good logical way of presenting this paradox because, with our present evolved intuition on motion, the Zeno paradox does not appeal to the modern common sense. But the essential idea is that if one subdivides the motion into smaller and smaller parts indefinitely, the distance  travelled at each instant approaches zero and then, there is no motion. This freezing of instantaneous position in motion is suggested as the freezing of the whole motion. But the argument is not valid for uniform motion since the time taken to move is proportionately smaller, and therefore the smallest divisions of space just takes the correspondingly smallest durations of time, with the same velocity, never freezing. 

However, consider some type of `motion' which starts off sluggishly and then speeds up.  So, in the initial durations there is hardly any movement and then the speed picks up and it starts moving uniformly, as in a traffic stop. If the traffic light stays green only for a short time, comparable to the time required to start moving even a short distance, the speed will never really pick up, and the time to move a given distance keeps increasing faster and faster, with a decreasing duration of the green signal. Finally when the light stays on for a duration smaller than the time to press the accelerator, the motion really freezes. This pseudo-Zeno effect happens because the motion is nonlinear in time and not uniform – starts off slowly and then picks up. This simple analogy, though not perfect, helps to understand the nature of the quantum Zeno effect. In quantum mechanics, a change of a state, or the quantum motion, is equivalent to a rotation of the state vector in an abstract space. A measurement is a projection of the vector along a chosen direction in its space. If the vector rotates by a very small amount and if a projection to its original direction (initial state) is taken, the length of the projection is almost the same as the length of the vector itself; the projection is different from the original vector only in the second order (square) in small quantities, and hence the difference is negligible.  So, if the system is not allowed to evolve for sufficient time before a measurement, that is equivalent to a small rotation and a projection, and the projection is almost the same as the original state. If there are frequent small rotations and projections (measurements) then the projections remain close to the original vector, and the state really does not evolve. It freezes in the original state, and this is the quantum Zeno effect. This can happen in any frequent quantum measurement. Sometimes this is paraphrased by saying that the watched kettle never boils. 

What is the relevance of the quantum Zeno effect for the Schr\"odinger’s cat? If there are frequent observations made on the system, does the cat live to its natural life expectation? (of course, the Zeno effect is not expected to prolong the natural life expectancy of a living being, though this is a theoretical possibility if quantum mechanics applies also to the live macroscopic state!) A more important question is to ask whether the quantum Zeno freezing occurs if the observation is done on just any part of the quantum system that would indicate the change of state, or whether it has to be on the initial member of the causal chain leading to the cat’s death. Can one delay the cat’s death by watching the cat continuously, or does one have to make observations on the quantum mechanical atom for the quantum Zeno freezing to occur?

This problem has not been analyzed to my knowledge. But we can make some immediate common sense statements. The causal chain involved in the evolution of the state of the Schr\"odinger cat is well defined and the starting point of the avalanche is the decay of the atom. If the cat dies of some other causes while in the box, that has no causal implication to the state of the atom. So, by looking at the state of the cat it is not logically and causally necessary to conclude that the atom has decayed, whereas if the atom is found decayed then the cat is most likely dead, if the apparatus is working as indented. So, we can safely come to the technical conclusion without going into the machinery of quantum mechanics that the quantum Zeno freezing of the possibility of the cat's death can happen only by the frequent observations on the state of the atom, and not by the frequent observations of the state of the cat. 

But this conclusion seems to depend on the existence of a definite causal chain with time delays between various events leading to the cat’s demise. A strictly quantum mechanical analysis using quantum states and their superposition on the other hand shows that the system never evolves into an entangled superposition from the initial state $\left|\psi\right>=\left|A+\right>\left|CL\right>$,  in which the atom has not decayed and the cat is alive, if repeated and frequent observations are made on just the cat. The cat can only `jump' from the state $\left|CL\right>$ to the state $\left|CD\right>$ without ever forming an entangled state with the atom! This contradiction with our earlier common sense conclusion happens because the quantum mechanical changes in the wavefunctions implied in the entangled state does not incorporate the causal time delays, and this is a deficiency of the way these states are written formally. This can be modified within the formalism of quantum mechanics and mutual consistency can be achieved in principle.

\section{The Quantum Muddle}
Karl Popper had drawn attention to a serious problem in the interpretation of quantum mechanics \citep{Popper-Schism}. He accused quantum physicists of being confused by the `quantum muddle'. Briefly stated, the quantum muddle refers to physically associating the wavefunction with the physical system, as a physical equivalence. The wave-particle duality is one such association often used – physicists refer to a particle having a wave property and the wavefunction is associated with this physical wave property of the particle. Popper argued that this is the source of most of confusions and paradoxes of the quantum theory. He reasoned that when one talks about a probability distribution function applicable to a population, one means that every statistical property of the population can be derived using the probability function. The formulae for the calculation resemble the ones used in quantum mechanics. But nobody, rightly, physically equates the probability distribution function with the individuals making up the population. Nobody would say that distribution function is a physical property of the individual. Seeing the individual with specific properties – a particular value for height, weight, hair colour etc., does not change the probability distribution function itself, and one would not talk about a `collapse of the function' when an observation of an individual results in a set of definite values of the properties of the individual.

The importance of Popper’s observation is the following: if a tight physical equivalence is not attributed to the correspondence between the wavefunction and the physical system, then the wavefunction is more of a tool for description. Even if a genuine ontology is associated with the wavefunction, if it has only a statistical role  then we can stop attributing the simultaneous existence of contradictory properties in the same physical object (when the wavefunction splits at a double-slit, the particle does not). While the wavefunction of the Schr\"odinger’s cat is in a superposition of two states corresponding to its being alive and being dead, the cat itself is not in a physical state of being dead and being alive. This is fully consistent with the mathematics of quantum mechanics and its predictions, and it takes away the confusion and perhaps the popular surprise from the often quoted and debated paradoxes. 

Einstein had emphasized a similar point about quantum mechanics, especially in his later writings. In the essay `Physics and Reality' \citep{Eins-Franklin},  Einstein wrote, 
\begin{quote}
	The $\psi$-function does not in
	any way describe a condition which could be that of a single system; it
	relates rather to many systems, to `an ensemble of systems' in the sense of
	statistical mechanics. If, except for certain special cases, the $\psi
	$-function furnishes only statistical data concerning measurable magnitudes,
	the reason lies not only in the fact that the operation of measuring
	introduces unknown elements, which can be grasped only statistically, but also
	because of the very fact that the $\psi$-function does not, in any sense,
	describe the condition of one single system. The Schr\"{o}dinger equation
	determines the time variations which are experienced by the ensemble of
	systems which may exist with or without external action on the single
	system.
\end{quote} 
I have argued elsewhere  that the extreme view taken by Popper -- that one should dissociate the single physical system (like a particular atom of an ensemble of atoms) from the wavefunction that describes the quantum behaviour of the ensemble -- is not consistent with quantum mechanical experiments. One can devise experimental situations from which one has to conclude that even at the single particle level `something' splits into two at the double-slit arrangement in the experiment. It is not the particle, and it does not carry energy. But it holds and encodes the properties of the physical system including information on its energy, momentum etc. What is its physical reality is a question that needs to be formulated and addressed in future. Answering this question will also clarify almost all contentious issues of interpretations of quantum mechanics \citep{Unnikrishnan-RQM}.\footnote{I found the definite answer only recently, more than a decade after writing this article, that the physical entity that superposes in quantum mechanics is the `action', the physical quantity introduced by W. R. Hamilton in 1833~\citep{Hamilton-S}. This also lead to completion of Hamilton's mechanics. A full account of the formalism that completely eliminates all foundational problems of quantum mechanics is now at hand, and it can be found in the reference \citep{Unnikrishnan-RQM}.}  

\section{The Boundary Between the Classical and the Quantum World}
One of the important issues that have been discussed extensively in the context of the Schr\"odinger’s cat is that of a boundary between the microscopic quantum world and the macroscopic classical world. Nobody knows whether there is such a boundary where the physicals laws are different. At present the understanding is that the physical laws are uniform across the two worlds, and they are the quantum mechanical rules, but the observed behaviour changes as one moves up to the microscopic world. Thus one does not see interferences of relatively large objects passing through small holes, and one does not see unexplainable randomness or correlations when one deals with macroscopic objects, whereas observations of microscopic atomic entities give results that are unfamiliar in the macroscopic world of common experience. 

The question arises, how large or massive an object can be described by a linear superposition of its physical states? Quantum mechanics would suggest that there is no limit and any object in this Universe, and perhaps the Universe itself, should be described by a wavefunction that could be in some complex superposition. But we also need to explain adequately why we do not see routinely the manifestations of a description in terms of superposition, like coherent oscillations between various possible states, drastic uncertainties and unpredictability in measurement results etc.
One of the important issues that is of relevance in the experiments on macroscopic systems is the need to isolate the system from its immediate environment or interactions of various kinds. Only then the quantum mechanical aspects become visible in the experiment. The loss of quantum mechanical superposition due to a large number of tiny interactions with the external environment is called `decoherence'. This is thought to be the fundamental reason why large objects familiar in the everyday world are not in any quantum mechanical superposition of states. Mathematically it is possible to show that the decoherence mechanisms do spoil \emph{coherent superpositions}. \emph{But the same mathematical analysis also implies that the system is still described by a superposition, albeit `incoherent'}. What this means is that the phase factor that appears in the addition of the two states fluctuates and does not have any fixed value, but it is still a superposition. For the case of the Schr\"odinger’s cat (`superposition version') an incoherent superposition results if the sign in between the states $\left|CL\right>$ and $\left|CD\right>$ fluctuates randomly between `$+$' and `$-$'. This still gives what would be seen classically – either a dead cat or a live cat, but the state of the individual cat is not a definite $\left|CL\right>$ or a definite $\left|CD\right>$. Of course, one can claim without inconsistency that in reality the state of affairs is like this and that all living and nonliving things in this world are in some incoherent quantum mechanical superposition till an observation is made to determine the state.

\section{The Schr\"odinger Cat and the Interpretations of Quantum Mechanics}
Before we end the discussion we will have a brief look at how the Schr\"odinger’s cat is viewed by different interpretations of quantum mechanics. The official view is that of the old Copenhagen interpretation in which the state of the cat is not entitled to a physical reality until an observation is made. (As an aside, the interpretation of the current theory of quantum mechanics was developed in G\"ottingen, by the school led by Max Born in which W. Heisenberg and P. Jordan were the early contributors, and not in Bohr's school in Copenhagen). In this view, the superposed or entangled wavefunctions allow a description of the state of the cat in an abstract sense, but do not imply anything further in terms of ontology. An observation is defined as a suitable interaction with a classical macroscopic device that reduced the state to a real and definite physical state. The view is entirely consistent, though unsatisfactory to many at the level of physical ontology and the physics and philosophy of (physical) reality.  

Another consistent and yet physically empty (in the sense of being beyond the grasp of empirical methods) view is that of the `Many-Worlds' interpretation of quantum physics. In this view, each component of the wavefunction is a physical reality in a Universe of its own, and therefore superposition signifies a totality of these definite states in different Universes. The evolution of the state is a coherent branching of the state, and of an observer himself, into different Universes,  everything being subject to the laws of quantum mechanics. `Naturally', some definite result is obtained in the measurement in each of the non-overlapping Universes. Thus the same cat is dead in some Universe, and alive in another, and there is no issue of the collapse of the wavefunction, nonlocality etc. Only, Schr\"odinger would have found the entire set up more ridiculous than his example of the superposed cat itself.  

An important interpretational view is that of L. de Broglie and D. Bohm, in which there is a hidden physical reality to the wavefunction and there is an observable physical reality to the particle \citep{Bohm-Hiley}. The associated quantum wave goes into a superposition of states, and the particle is guided by the superposed wave in a complicated trajectory. This view is also consistent for single particle quantum mechanics, whereas the many questions that arise in multi-particle situations have not been not been answered satisfactorily. I feel that the de Broglie-Bohm view runs into some problems handling the Schr\"odinger’s cat. In the example of the double-slit, the particle passes through one of the slits and propagates to the screen guided by the wavefunction. But in each trial of the experiment employing particles one by one, the particle passes either through the upper slit or through the lower slit, and generally remains in the upper or lower part of the screen. If one translates this to the case of the cat, the state of being dead or alive materializes during the evolution itself, and any quantum mechanically specific property like interference should come from the interference of the guiding waves. In the machinery of quantum mechanics it is possible to retrace the evolution of the wavefunction without making any interaction or observation during the evolution and such a reversible evolution should be accountable in any valid formalism. The Copenhagen view has no problems in this regard, but the de Broglie-Bohm view seems to face some problems because once the cat is dead in a physically real sense, there is no possibility of reviving it even with the magic of quantum mysteries. Therefore, any interpretation of quantum mechanics that ascribes physical reality to the states before an observation is made will run into severe problems and inconsistency when it has to handle the Schr\"odinger’s cat (the Many-Worlds interpretation passes this test because the reality is branched into different Universes). On the other hand, even the Schr\"odinger’s cat can be brought consistently within the Bohmian quantum mechanics if we treat the state of death as a mere physical rearrangement of a set of particles and molecules obeying the laws of quantum mechanics. Then the evolution from being live to being dead is a quantum mechanical evolution that is in principle reversible, and can be treated consistently within all these interpretations. This aspect needs further study and thought.  

\section{Can the Cat Teach Us Further?}
The Schr\"odinger’s cat has played a significant role in making us think about several issues related to the interpretation of quantum mechanics and its applicability at various scales of physical phenomena. It has also motivated some challenging experimental activity in which people try to observe phenomena that are specific to quantum mechanics, in progressively larger and larger physical systems. At present there seems to be some clarity and much confusion on many of the issues, and all these can be traced to our lack of understanding of the meaning and ontology of the wavefunction. It is only when we understand the nature of the wavefunction, that we will be able to talk with clarity about superposition, entanglement, collapse, measurement etc.\footnote{A complete understanding of the wavefunction and the Schr\"odinger equation was achieved recently \citep{Unnikrishnan-RQM}. See the  section, 'Epilogue'.} Without doubt, the Schr\"odinger cat will remain alive, provoking and motivating interesting ideas, experiments and debates in the foundational aspects of the most challenging theory of our times. 

\section{Epilogue}
This article was written 15 years ago, for a volume edited by P. K. Sengupta in the remarkable series `History of Science and Philosophy of Science', of the Centre for Studies in Civilization, India. No changes in the descriptive part of the article seem to be required even now. However, a definite  advance happened quite recently, which redraws the entire canvas of the understanding of quantum mechanics, clarifying and resolving all the foundational issues, including the concerns expressed in examples like the Schr\"odinger cat. This consists of two breakthroughs about the correct understanding of both the mathematical structure and the physical content of quantum mechanics~\citep{Unnikrishnan-RQM}: 

\begin{enumerate}
	\item I have been able to prove in detail the hunch expressed by Einstein and Popper, that the Schr\"odinger equation and its wavefunction describe an ensemble of system, and not a single quantum system. In fact, the Schr\"odinger equation is readily shown to be the continuity equation for the probability density, $\partial \rho/\partial t=\nabla\cdot (\rho v)$, written in terms of its complex single valued square root, $\psi(t)=\sqrt{\rho}exp(i\phi)$.  Then the Born relation $\rho=\psi \psi^*$ is an exact mathematical relation, and not an additional postulate. The quantity $\phi(t)$ enables the representation of both $\rho$ and the dynamical velocity $v$ in the same function $\psi$. The $\psi$-function pertains to the ensemble, and it is a hybrid quantity consisting of the (real) square root of the ensemble probability density and the ensemble averaged complex valued periodic function of action. Then, it turns out that the superposition and the entanglement do not involve the physical states of the particle (or the cat); what superposes is merely the action function with a periodic property. 

\item The second result addresses the question, what then is the evolution equation for quantum dynamics if the Schr\"odinger equation is factually the continuity equation? The single particle dynamical equation is a modified Hamilton's equation, simply written as $\partial \zeta(S) /\partial t=i\hbar H \zeta$, where $\zeta(S)=exp(iS/\hbar)$ and $H$ is the Hamiltonian. This is universally applicable at all scales of masses, sizes and velocities. There is no fundamental distinction between the dynamics of microscopic and macroscopic bodies, and there is no quantum-classical divide. The entire probability space from zero to one is mapped in the values of the action spanning the small value $\hbar$, because of the periodicity of the function $\zeta(S)$. This new equation implies that the fundamental uncertainty relation is in fact $\Delta S\geq \hbar$.
\end{enumerate}
Together, these two results completely eliminate all the foundational and interpretational issues of quantum mechanics, while reproducing all the results involving quantum interference and correlation.

Thus, the alleged wave-particle duality of a particle is not the physical fact. A particle has no wave nature and there is no `matter-wave'; instead, the action of its dynamics manifests as a periodic complex valued entity, which defines the multiple possibilities of observables and their superposition. Since this `action-wave' carries action, and not energy or momentum, there is no collapse of the quantum state when observations are made. Just as a probability density of an ensemble is not altered or perturbed when an individual statistical event is realized, the wavefunction of the Schr\"odinger equation does not collapse when an observation is realized. The same remarks apply for an entangled wavefunction as well. The collapse of the wavefunction is eliminated from the quantum mechanics, and with it, the quantum measurement problem is solved entirely.

For the example of the Schr\"odinger cat, any single radioactive atom and the individual cat are always in their definite states, \emph{either} $\left|A+\right>\left|CL\right>$ \emph{or} $\left|A-\right>\left|CD\right>$, and never in a  superposition or entanglement. What is superposed are the relevant complex exponential functions of the action corresponding to the two systems. A single atom-cat pair is always in a definite joint state that will eventually jump from the initial state $\left|A+\right>\left|CL\right>$ to the fateful state $\left|A-\right>\left|CD\right>$, whether or not one is observing either system. What the Schr\"odinger equation and the evolving wavefunction describe is the situation in the statistical ensemble, consisting of large number of atom-cat pairs.  This definitely and completely answers all the questions raised in the context of the Schr\"odinger cat. All the ridiculous aspects disappear. The conscious observer and his friend play no role in the dynamical evolution. 

The details of this universal picture of clarity about quantum mechanics, the Schr\"odinger equation, and its wavefunction are described in the articles \citep{Unnikrishnan-RQM} and \citep{Unnikrishnan-GRVol}. 

\section*{Acknowledgements}
\noindent Some of the thoughts presented in this article evolved in the presence of `Jaldi', the playful cat in Paris (still very much alive, which is very reassuring). I sincerely thank Prof. Pradip Sengupta for skillfully provoking me to write the original article, about 16 years ago, and for thus helping me to acquire some important insights. Martine Armand helped in structuring the updated sections.

\end{document}